\documentclass[osajnl, twocolumn, superscriptaddress, 10pt]{revtex4-1} 

\usepackage{amsmath,amssymb,graphicx}
\pdfoutput=1
\begin{document}

\title{Waveguide couplers for ferroelectric optical resonators}

\author{Ivan S. Grudinin}

\affiliation{Jet Propulsion Laboratory, California Institute of Technology, 4800 Oak Grove Dr., Pasadena, CA 91109, USA}\email{Corresponding author: grudinin@jpl.nasa.gov}

\author{A. Kozhanov}
\affiliation{Center for Nano-Optics, Georgia State University, 29 Peachtree Center Ave SE, Atlanta, GA 30303, USA}
\author{N. Yu}
\affiliation{Jet Propulsion Laboratory, California Institute of Technology, 4800 Oak Grove Dr., Pasadena, CA 91109, USA}\email{Corresponding author: grudinin@jpl.nasa.gov}

\begin{abstract}We report a study of using the same material to fabricate a whispering gallery mode resonator and a coupler. Coupling to high Q whispering gallery modes of the lithium niobate resonator is demonstrated by means of the titanium-doped waveguide. The waveguide coupling approach opens possibilities for simpler and wider practical usage of whispering gallery mode resonators and their integration into optical devices. Copyright 2014. All rights reserved. 
\end{abstract}
\maketitle

Crystalline whispering gallery mode (WGM) resonators are useful in a growing number of applications\cite{1,2} including nonlinear optics\cite{3}, cavity optomechanics\cite{4}, microwave\cite{5} and THz photonics\cite{6}. Ferroelectric crystal materials such as lithium niobate (LN) and lithium tantalite (LT) are particularly attractive for WGM resonators because they are highly transparent in the infrared and visible ranges which results in Q factors of WGMs of above 100 million, have strong electro-optic coefficients which allow for electronic tuning of modes, and large nonlinear susceptibility for efficient frequency conversion and parametric processes. 

High indices of refraction of ferroelectric materials make evanescent-wave coupling by means of standard prism coupling techniques challenging. To achieve efficient coupling with the WGMs of a resonator in general, the wave vectors of the evanescent field provided by the coupler and that of the resonator mode should be matched. In addition, there should be a significant overlap of the two fields\cite{7}. In most coupling approaches this leads to a requirement for the refractive index of the coupler material to be higher than that of the resonator. The LN and LT have refractive index of over 2 in the near-IR. No optical fiber is available with such high refractive index to implement angle-polished\cite{8} or tapered fiber\cite{9} coupling mechanism. For the same reason the polymer waveguides\cite{10} are not suitable. Thus, the only practically used method to excite optical modes efficiently in such resonators relies on a diamond or rutile (TiO$_2$) prism having refractive indices higher than those of LN and LT, in combination with a carefully designed collimating optics. Nearly perfect coupling has been demonstrated with this approach\cite{6}. However, the diamond and rutile prisms are expensive and, in addition, the rutile has significant birefringence making the coupling procedure more complex. Waveguide coupling would not only enable more applications of high-index ferroelectric resonators, but also make it easier to work with lower index resonators due to simplicity of the approach.

Here we report a proof-of-principle demonstration for a LN WGM resonator coupling using a titanium-diffused waveguide fabricated on a wafer having the same index of refraction as the WGM resonator.

The LN waveguide was fabricated as follows. A 6 $\mu$m wide and 100 nm thick Ti stripe was lithographically defined along the Y crystalline direction on top of a polished X cut LN substrate using lift-off technique (Fig.1). Following the standard procedures\cite{11,12} the sample was then annealed at 1000 C in oxygen atmosphere for 10 hours to allow the metal to diffuse into the LN substrate and form a waveguide with sufficient refractive index contrast. 
\begin{figure}[htbp]
\centerline{\includegraphics[width=1\columnwidth]{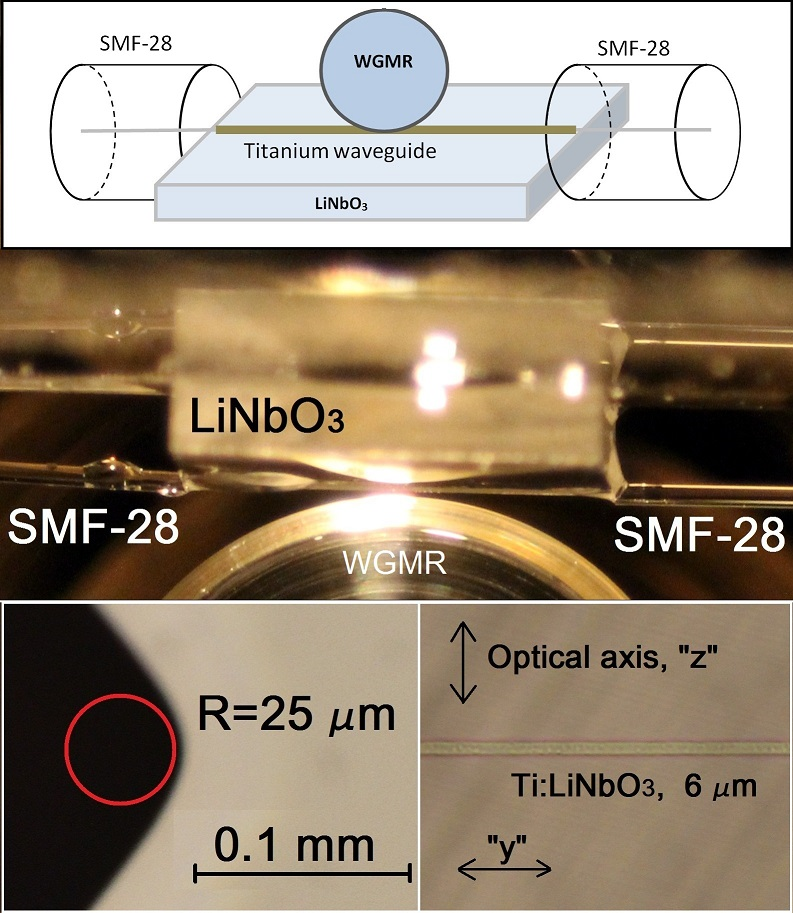}}
\label{fig:collage}
\caption{Coupling schematics (top). Photograph of the coupler and the WGMR (middle). Resonator edge and the waveguide photograph (bottom).}
\end{figure}
The WGM resonator with 3 mm diameter and 25 micrometers edge radius was fabricated from a z--cut LN wafer. The waveguide coupler and the resonator are arranged for edge coupling, i.e. the coupling is through the edge of the resonator as shown in Figure 1. In this arrangement, the TE waveguide mode experiences the ordinary refractive index and can be used to couple to the TM modes of a Z cut WGM resonator having electric field in the radial direction. On the other hand, the TM waveguide modes experience the extraordinary index and can be coupled to the TE modes of a Z cut resonator with electric field polarized along the optical axis transversely to the resonator surface. The waveguide substrate was cut, polished and glued to the cleaved single mode Corning SMF--28 fiber tips. The resulting fiber pigtailed waveguide device is used for lateral coupling to a Z--cut LN WGM resonator as shown in Fig. 1. While nearly perfect (1 dB loss) end-firing is possible for this configuration\cite{13} we observed 10\% transmission for the TE waveguide mode through the waveguide. This low transmission figure is explained by the poor matching of modes of our waveguide and the mode of the fiber, as well as by a large refractive index mismatch. Approximately 1\% transmission was observed for the TM polarization due to the TM field coupling with the TE slab modes\cite{14}. 
\begin{figure}[htbp]
\centerline{\includegraphics[width=1\columnwidth]{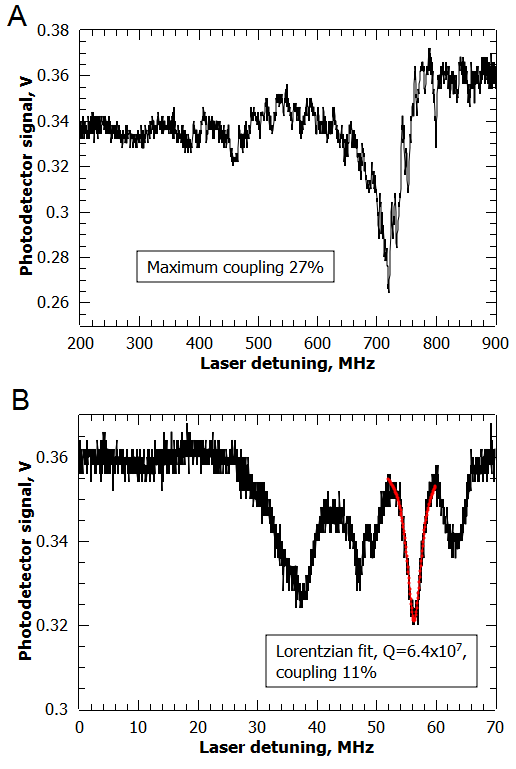}}
\label{fig:12}
\caption{A) Coupling to a number of TM WGMs is observed. B) Coupling to a high Q factor TM WGM.}
\end{figure}
With the experimental setup in Figure \ref{fig:collage}, we observed that coupling to a large number of different WG modes takes place upon placing the resonator edge to within 1 micron of the waveguide. The resonant dips have different quality factors and coupling efficiencies as shown in Fig. 2A and may be associated with WGMs having different number of field maxima in azimuthal and radial direction (mode indices). For example, in Fig. 2A a low Q WGM having up to 27\% coupling efficiency is visible along with a number of high Q WGMs having weaker coupling efficiency. In another portion of the spectrum coupling to high Q WGMs was observed as shown in Fig. 2B. A Lorentzian fit of one of the observed modes yields an optical Q factor of over 60 million with coupling efficiency of 11\%.

In addition to meeting the index requirement of the general evanescent wave coupling scheme, the waveguide based coupling scheme demonstrated here should be more advantageous than the prism-collimator approach for simplicity of integration. While non--optimized waveguides have high transmission loss and poor coupling efficiency\cite{conti}, other waveguide geometries may be used to reduce TM mode attenuation such as ridge waveguides\cite{14}. However, further studies are required to optimize the optical and geometrical design parameters of the titanium-diffusion waveguides for reduced transmission loss and increased coupling efficiency. It should be pointed out that the titanium waveguide approach can also be configured for planar coupling to ferroelectric resonators similarly to the polymer waveguides\cite{10} and for on-chip laterally coupled nonlinear optical devices.

This work was carried out at the Jet Propulsion Laboratory, California Institute of Technology under a contract with the National Aeronautics and Space Administration, with support from NASA Center Innovation Fund and JPL Research and Technology Development Program.

\end{document}